\documentclass[fleqn,10pt]{wlscirep}
\usepackage[utf8]{inputenc}
\usepackage[T1]{fontenc}
\usepackage{xcolor}
\usepackage{lineno}

\title{Landscape-induced spatial oscillations in population dynamics}

\author[1]{Vivian Dornelas}
\author[2,3,4]{Eduardo H. Colombo}
\author[2]{Crist\'obal L\'opez}
\author[2]{Emilio Hern\'andez-Garc\'{\i}a}
\author[1,5,*]{Celia Anteneodo}
\affil[1]{Department of Physics, PUC-Rio, Rua Marqu\^es de S\~ao Vicente, 225, 22451-900, Rio de Janeiro, Brazil.}
\affil[2]{IFISC (CSIC-UIB), Campus Universitat Illes Balears, 07122, Palma de Mallorca, Spain.}
\affil[3]{Department of Ecology and Evolutionary Biology, Princeton University,  Princeton NJ 08544, USA.}
\affil[4]{Department of Ecology, Evolution, and Natural Resources, Rutgers University, New Brunswick, NJ 08901, USA.}
\affil[5]{Institute of Science and Technology for Complex Systems, Rio de Janeiro, Brazil.}
\affil[*]{celia.fis@puc-rio.br}

\begin{abstract}
We study the effect that disturbances in the ecological landscape exert 
on the spatial distribution of a population 
that evolves according to the nonlocal FKPP equation. 
Using both numerical and analytical techniques, we characterize, as a function of the interaction kernel, the three 
types of stationary profiles that can develop near 
abrupt spatial variations in the environmental conditions vital for population growth: 
sustained oscillations, decaying oscillations and exponential relaxation towards a flat profile. 
Through the mapping between the features of the induced wrinkles and the shape of the interaction kernel, we discuss how heterogeneities can reveal information that would be hidden in a flat landscape.  
\end{abstract}
\begin{document}

\flushbottom
\maketitle

\thispagestyle{empty}

\section{Introduction}
\label{sec:intro}

The Fisher-Kolmogorov-Petrovskii-Piskunov (FKPP) equation \cite{fisher1937, Kolmogorov, Cencini2003}
is the standard model describing, at a continuum level,
the spatio-temporal dynamics of a population of individuals that diffuse, grow and compete for resources. 
In one dimension, it is given by
\begin{equation}
\partial_t \rho (x,t) = D\partial_{xx} \rho(x,t) + a\rho(x,t)-b\rho^2(x,t)\, , 
\label{eq:fkpp}
\end{equation}
where $\rho(x,t)$ is the population density at position $x$ and time $t$, 
$D$ is the diffusion coefficient, $a$ is the (clonal) reproduction rate, 
and $b$ is the strength of (intraspecific) competition that bounds population growth. 

In Eq.~(\ref{eq:fkpp})  competition is local,  in the sense that it occurs at scales 
much smaller than those associated with the diffusion process. 
However, competition processes might also extend to larger scales. 
This can be promoted by the underlying dynamics of interaction mediators (e.g., shared resources),
such that even if individuals' actions are  locally initiated, the effects of these actions propagate to the surroundings. 
Along the lines of the Turing mechanism~\cite{turing}, the mediator dynamics can be explicitly modeled using an additional reaction-diffusion equation, an approach that has been widely applied to water-vegetation systems~\cite{klausmeier,hardenberg2001,eppinga2008regular,rietkerk,hillerislambers}. 
Also in the context of vegetation dynamics, competition among plants can be mediated by roots~\cite{ricardogrl,oto2013}, that extend beyond the surface vegetation.  
More generally, spatially-extended interactions between many types of organisms can be generated by various other mechanisms, such as
acoustic communication~\cite{ricardoprl},  
exchange of physico-chemical signals~\cite{kiorboe2011,ojalvo2017,bauerle2018,potts2019spatial}, 
and spatial exploitation, as present in the case of sensile~\cite{connel} and territorial~\cite{carter2015} organisms.

\begin{figure}[h!]
\begin{center}
\includegraphics[width=0.6\columnwidth]{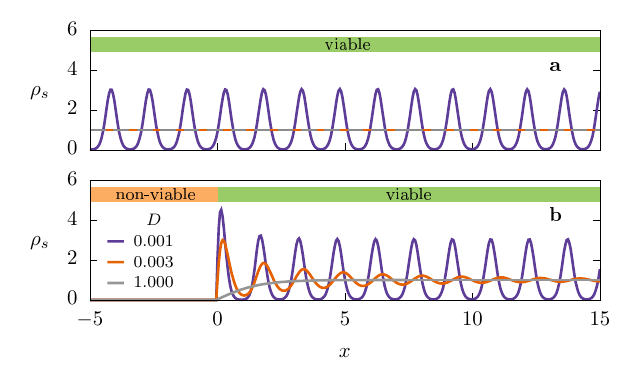}
\end{center}
\caption{
  {\bf Population distribution in a medium which is
(a) homogeneously viable, 
(b) heterogeneous, with viable and non-viable regions.}
Depending on the values of the parameters in Eq.~(\ref{eq:model}),  
spatial patterns can develop around the uniform steady state in 
(a), and they are preserved in the viable region of the corresponding case in (b). 
But even when the steady state is uniform in (a),  decaying oscillations can emerge in (b). 
Parameters are $a=b=1$, values of $D$ are given in the legend, and for kernel $\gamma_q$ 
defined in Eq.~(\ref{eq:kernel}),  we fix  $q=-0.5$ and $\ell=2$. 
For panel b, $A$ in Eq.~(\ref{eq:semiinfinite}) is $A \to \infty$.
}
\label{fig:profiles}
\end{figure}

Although the intraspecific competitive interaction is
established by other species or substances that act as mediators,  when their  timescales are much shorter than the population ones, a single equation for the distribution of individuals can be derived~\cite{ricardo2014}. 
This effective equation contains a nonlocal term describing the influence of individuals at a distance. 
Due to the often complex web of processes regulating the mediator dynamics, a useful phenomenological approach is to incorporate their effects with an interaction kernel (also called influence function) $\gamma(x)$, describing how the effective interaction between individuals decreases with their
distance $x$.
Furthermore it allows to address rather generally the impact of distance-dependent competition regardless of the mechanisms behind it.
Then, Eq.~(\ref{eq:fkpp}) is extended as~\cite{Sasaki1997,Fuentes,pre2004}  
%
\begin{equation}
\partial_t \rho(x,t) = D\partial_{xx}\rho(x,t) + a\rho(x,t) - 
b \rho(x,t)[\gamma \star \rho](x,t) \,,
\label{nonlocalfkpp}
\end{equation}
where $[\gamma \star \rho](x,t) \equiv  \int_{-\infty}^{\infty}\gamma(x-x')\rho(x',dt) dx'$, 
and $\int_{-\infty}^{\infty}\gamma(x)dx=1$. The particular shape and characteristic scales of $\gamma$ effectively embody the details of the interaction mechanisms.
At a difference from the original FKPP, 
the nonlocal FKPP equation, given by Eq.~(\ref{nonlocalfkpp}), can exhibit self-organized structures (as depicted in Fig.~\ref{fig:profiles}a) 
depending primarily on the particular properties of the kernel 
and, secondarily, on the values of the diffusion and reproduction rates~\cite{Fuentes,pre2004,physicaD,pigolotti}. 
This is a minimal continuous-field description of population of individuals that compete nonlocally, 
containing the essential ingredients used to model diverse species dynamics~\cite{pre2004}.
 
Moreover, Eq.~(\ref{nonlocalfkpp}) assumes a homogeneous environment, 
which is implicit in the constant coefficients. 
But, actually, in biological systems, 
environmental factors suffer spatial variations~\cite{berti,perry2005,landscapeBook}.  
In this paper, we exploit that they can stress the system and resonate with the internal scales~\cite{Taylor2020,krause2020,gaffney2019,garcia2012noise,Sasaki1997},  
generating spatial oscillations in the distribution of the population  
that can serve to unveil hidden information. 
In order to do that, we consider the following extension of Eq.~(\ref{nonlocalfkpp}),
\begin{equation}
\partial_t \rho(x,t) = D\partial_{xx}\rho(x,t) + \Psi(x)\rho(x,t) - 
b \rho(x,t)[\gamma \star \rho](x,t)\,, 
\label{eq:model}
\end{equation}
where the spatially-dependent reproduction rate, $\Psi(x)$, reflects the overall habitat quality 
at a given location $x$~\cite{landscapeBook}.

Particular forms of $\Psi$, accounting for diverse complex spatiotemporal features of natural environments,  have been considered in previous studies~\cite{colombo2015,silva2014,Sasaki1997,page2003}. They have shown how this spatial dependence can modify the stability domains or even generate new states that were absent otherwise. For the particular case in which environmental disturbances are random, 
these results can be framed in the context of \emph{noise-induced transitions}\cite{horsthemke1984noise,garcia2012noise,ridolfiBook}.

In this work, we focus on sharp changes in the spatial  environmental conditions  
relevant for the population under consideration~\cite{landscapeBook,perry2005,fonseca2013}. 
This kind of change is found in diverse situations in nature, e.g., 
at the interface between forest and grassland~\cite{landscapeBook},  
at the bounds of oases~\cite{berti} or harmful regions~\cite{tarnita2017},  
or in artificial lab experiments~\cite{perry2005}, 
where there is a neat contrast of  spatial domains with different growth rates. 
Contemplating these cases  justifies attributing the Heaviside step 
or rectangular functions to $\Psi(x)$. 
For Turing-like models, it has been shown  that the existence of a step-function hetereogenity can promote the formation of decaying oscillations even when the system is stable under homogeneous conditions~\cite{page2003}. We extend this discussion by noting that there is a close relation between the landscape-induced states and the underlying dynamics of the interaction mediators which,  in our case,  is captured by the influence function.

We perform a systematic exploration of the model parameter space and investigate the emergence of the three kinds of stationary (long time) population profiles that can develop   from the interface between regions of contrasting characteristics: 
{\it sustained oscillations} (or spatial patterns, without amplitude decay), 
{\it decaying oscillations} (with decreasing amplitude from the interface)
or {\it exponential decay} towards a flat profile. 
These behaviors are schematically depicted in Fig.~\ref{fig:profiles}b.
Ultimately, we report the existence of a one-to-one mapping between the influence function parameters and the oscillations features, allowing us to extract details of the interaction from the pattern images~\cite{zhao2020}.

The paper is organized as follows.  
In Sec. ~\ref{sec:preliminaries} ,
we provide introductory information with general considerations  
about the homogeneous environment as a frame of reference.  
In Sec. ~\ref{sec:heterog},
we present the main results for 1D landscapes with sharp changes. 
Additionally, outcomes for 2D landscapes are 
displayed.  
In Sec. ~\ref{sec:inferring},  we discuss how information about the interaction kernel can be extracted from 
observable oscillations. 
A summary of the main findings and discussion are presented in Sec.~\ref{sec:final}.
%

\section{Preliminaries}  
\label{sec:preliminaries}

In this section, we first define the main class of influence functions 
that will be used in numerical examples throughout the paper. 
We also revisit the linear response analysis for the homogeneous environment, 
which serves as a reference frame for the more complex heterogeneous case.

\subsection{Interaction kernel}
\label{sec:interact}

\begin{figure}[h!]
\begin{center}
\includegraphics[width=\columnwidth]{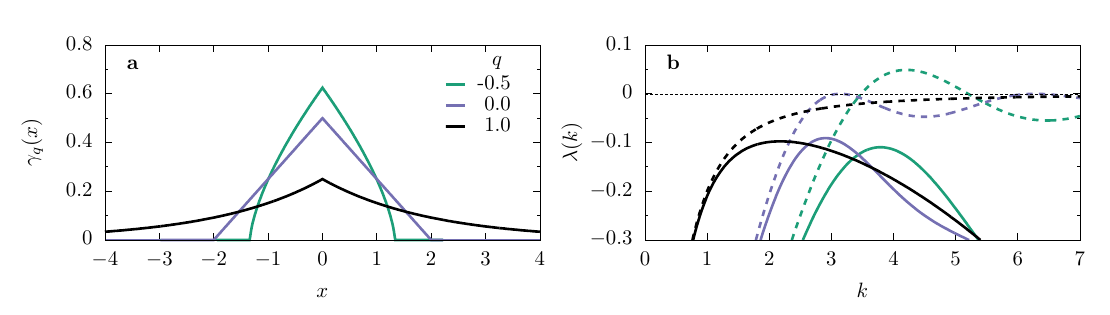}
\end{center}
\caption{ {\bf Interaction kernel and mode stability in a homogeneous medium. } 
(a) $\gamma_q(x)$,  defined in Eq.~(\ref{eq:kernel}), for the values of $q$ 
indicated on the figure, and $\ell=2$.
(b) Mode growth rate $\lambda(k)$ given by Eq.~(\ref{eq:dispersion}), for $a=b=1$,  
with $D=0$ (dashed lines) and $D=0.01$ (solid lines), corresponding to 
the values of $q$ plotted in (a). The case $q=0$ (triangular kernel) is the critical one, 
for which the maximal value of $\lambda(k)$ at finite $k$ is zero when $D=0$. Notice that, when diffusion 
is absent, the mode growth rate is proportional to the kernel Fourier transform (see Eq.~(\ref{eq:dispersion})).
}
\label{fig:kernel}
\end{figure}

We have chosen a family of influence functions that 
allows us to continuously vary its compactness:  
\begin{equation}
\gamma_q(x) = N_q [1 - (1-q)|x|/\ell]_+^{1/(1-q)} \,\equiv\,  N_q \exp_q(|x|/\ell) \, ,
\label{eq:kernel}
\end{equation} 
where  $q$ and $\ell$ control the shape and scale of the kernel, respectively,  and $N_q$ is a normalization constant. 
The subindex + means  $[z]_+=z$, if $z > 0$, and $[z]_+=0$ otherwise.
This kernel is based on a generalization of the exponential function, 
known as $q$-exponential~\cite{tsallis2009}. 
In the limit $q\to 1$, the standard exponential is approached  yielding 
$\gamma_1(x) \propto e^{-|x|/\ell}$.  
 The kernel shapes, for different values of $q$ are illustrated in  Fig.~\ref{fig:kernel}a. 
As we will see, it is specially relevant the fact that, 
only for $q<0$, the Fourier transform of $\gamma_q(x)$ can take negative values. 
Then, we focus on the range $-1 \le q <1$, around this critical value.
Moreover,  in this range, the interaction is restricted to a finite region and 
the kernel moments are well-defined, a fact that will facilitate both the numerical 
and theoretical approaches. 
 The family of stretched exponential kernels was also considered for comparison (see {\it Supplementary Information}).

\subsection{Homogeneous landscapes}
\label{sec:homog}

For a homogeneous landscape,  with $\Psi(x)=a$, 
the linearization of Eq.~(\ref{eq:model})  around its uniform solution  $\rho_0=a/b$,  done
by setting 
 $\rho(x,t) = \rho_0 + \varepsilon(x,t)$ (with $\varepsilon/\rho_0 \ll 1$),  gives
$\partial_t\tilde{\varepsilon}(k,t)= [-Dk^2 - a\tilde{\gamma}(k)]\tilde{\varepsilon}(k,t)$ in Fourier space, 
where $\tilde{\gamma}(k) = \int_{-\infty}^{\infty} \gamma(x) e^{- i k x} dx$ 
is the Fourier transform of the interaction kernel $\gamma$.
The factor between square brackets represents the  growth rate of mode $k$,
\begin{equation}
\lambda(k) = -Dk^2 - a\tilde{\gamma}(k)\, ,
\label{eq:dispersion}
\end{equation}
which, for the considered kernels, is a real function, and whose  shape is plotted in Fig.~\ref{fig:kernel}b, for each kernel $\gamma_q$  
shown in Fig.~\ref{fig:kernel}a. 
It is the important quantity that will appear all throughout the paper, since solutions of the transformed linearized equation satisfy $\tilde\varepsilon(k,t)= \tilde\varepsilon(k,0) {\rm e}^{\lambda(k)t}$.
Thus, if $\lambda(k)<0$ for all $k$, any initial perturbation will fade out, such that in the long-time limit the population distribution, $\rho(x)$, will be flat. 
On the contrary, if there are unstable modes, with $\lambda(k)>0$, stationary spatial oscillations will be produced 
with a characteristic mode $k^\star$ (the maximum of $\lambda$), which is 
the initially fastest growing one~\cite{colombo12}.

From Eq.~(\ref{eq:dispersion}), $\lambda(k)>0$ occurs for sufficiently small $D$  
if the Fourier transform of the kernel takes some negative values. 
Then, by substitution of $\tilde{\gamma}_q$ into Eq.~(\ref{eq:dispersion}), 
we conclude that sustained oscillations can only appear if $\gamma_q(x)$ 
is sub-triangular, i.e., $q<0$ (in the critical case $q=0$, $\gamma_q(x)$ 
produces the triangular kernel, whose Fourier transform is 
$\tilde\gamma_0(k)= \sin^2(k \ell)/( k \ell)^2$).
This is a necessary but not sufficient 
condition that arises by imposing $\lambda(k^\star)>0$ in the most 
favorable case $D=0$ (hence $\tilde{\gamma}(k^\star)<0$),  
to induce the growth of certain modes. 
In contrast, for  $q \geq 0$,  
the uniform state is intrinsically stable (that is, independently of the remaining parameters).
In Fig.~\ref{fig:kernel}b, we plot the mode growth rate for $D=0$ and $D>0$, 
which shows  how diffusion affects mode stability, damping inhomogeneities in the population distribution.

Concerning the interaction length $\ell>0$, 
when $D=0$, it simply scales the wavenumber as $k\ell$. 
Therefore, when $\ell$ goes to zero (implying local dynamics), $\lambda(k) \to \lambda(0)<0$, 
meaning that patterns go continuously to a flat profile in that limit.
In contrast, for $D>0$, the first term in Eq.~(\ref{eq:dispersion}) 
has a more homogenizing effect the larger is $k^\star$, hence the smaller is $\ell$. As a consequence, 
despite interactions are nonlocal, patterns emerge only 
for $\ell$ above a critical value~\cite{colombo12}. 
For $D>0$, there is also a critical reproduction rate, $a_c$, such that  sustained oscillations 
emerge only for $a>a_c$.

In summary, in the cases where $\lambda(k^\star) \le 0$, i.e., either $q \ge 0$, or $q < 0$ with 
sufficiently large $D$ (or, alternatively, small enough $\ell$ or $a$),   
information regarding the interaction scale $\ell$ 
or other details of the kernel profile  are not stamped in the spatial distribution $\rho(x,t)$, which becomes 
uniform at long times.

\section{Heterogeneous landscapes}
\label{sec:heterog}

In this section, the heterogeneity of the landscape is introduced by assuming that its profile can be written as  
$\Psi(x) = a + \psi(x)$, where $\psi(x)$ represents the spatial variations 
of the environment around a reference level $a$.

The results that we will present were obtained through theoretical and numerical techniques. 
The theoretical approach is based on the mode linear stability analysis discussed in the 
previous section. 
Numerical integration of  Eq.~(\ref{eq:model}), 
starting from a homogeneous state plus a random perturbation,  was performed following an explicit forward-time-centered-space scheme, 
with boundary conditions suitably chosen for each case (see {\it Supplementary Information} for details).

\subsection{Refuge}

As a paradigm of a heterogeneous environment with sharp borders, we first  consider  that 
the spatial variations around the reference level $a$ are given by
\begin{equation} 
\label{eq:refuge}
\psi(x) =   - A[1- \Theta(L/2 -|x|)]  \,, 
\end{equation}  
where $\Theta$ is the Heaviside step function. 
With $A>0$, it represents a refuge  of size $L$ with growth rate $a$ 
immersed in a less viable environment with growth rate $a-A$.
In a laboratory situation, this can be constructed by means of  a mask delimiting a region 
that protects organisms from some harmful agent, for instance, shielding bacteria from 
UV radiation~\cite{perry2005}.   
In natural environments, this type of localized disturbance appears due to changes in 
the geographical and local climate conditions\cite{landscapeBook}, 
or even engineered by other species~\cite{tarnita2017}.

\begin{figure}[h!]
\begin{center}
\includegraphics[width=0.5\columnwidth]{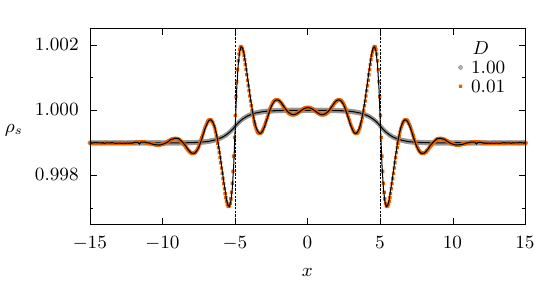}
\end{center}
\vspace{-0.2cm}
\caption{
{\bf 
Stationary population density $\rho_s$ vs. $x$ in a refuge.}
This heterogeneous environment is defined by Eq.~(\ref{eq:refuge}), 
with $a=b=1$, $A=10^{-3}$ and $L=10$. 
The vertical lines indicate the refuge boundaries.
We used the kernel $\gamma_q(x)$,  with $q=0.1$ and $\ell=2$, 
and two different values of $D$. 
Symbols are results from numerical integration of Eq.~(\ref{eq:model}) under periodic boundary conditions,  
and solid lines from the small-$A$ approximation 
given by Eq.~(\ref{eq:rhoS}), 
in excellent agreement with the exact numerical solution. 
Recall that, in a homogeneous environment, no oscillations  appear for $q \ge 0$.
} 
\label{fig:revealed}
\end{figure}

In Sec.~\ref{sec:homog},
we have seen that the uniform distribution is intrinsically stable when $q \ge 0$.  
In contrast, when there are heterogeneities in $\Psi(x)$,
spatial structures can emerge even if $q\ge 0$, 
as illustrated in Fig.~\ref{fig:revealed} for the case $D=0.01$.

In the limit of weak heterogeneity, i.e., 
under the condition $|\psi(x)|/a \ll 1$, we obtain an approximate analytical solution 
assuming that the steady solution of  Eq.~(\ref{eq:model}) can be expressed in terms 
of a small deviation  $\varepsilon_s(x)$ around the homogeneous state $\rho_0=a/b$. 
Then, we substitute  $\rho_s(x)=\rho_0+\varepsilon_s(x)$ into the stationary form of 
Eq.~(\ref{eq:model}), discard terms of order equal or higher than 
${\cal O}(\varepsilon^2, A\varepsilon,A^2)$, and Fourier transform, obtaining
\begin{equation}
\label{eq:eqepsilonFT}
  \tilde\varepsilon_s(k) = \dfrac{ \rho_0 \tilde{\psi}(k)}{-\lambda(k)}\,,
\end{equation}
where $\lambda(k)$ was already defined in Eq.~(\ref{eq:dispersion}) 
and $\tilde{\psi}(k)$ is the Fourier transform of the small fluctuations 
in the landscape quality, which for the case of Eq.~(\ref{eq:refuge}) is
$\tilde\psi(k)=  A[2\sin(Lk/2)/k -2\pi \delta(k)]$. 

Finally, assuming that $\lambda(k^\star)<0$, 
the  steady density distribution is  given by
\begin{equation}  
\rho_s(x) \,=\, \rho_0  +\varepsilon_s(x)  \,=\, \rho_0  +  
{\cal F}^{-1}\Bigl(\dfrac{ \rho_0 \tilde{\psi}(k)}{-\lambda(k)}\Bigr) \,,
\label{eq:rhoS}
\end{equation}
where the inverse Fourier transform ${\cal F}^{-1}$ must be numerically computed in general. 
For small heterogeneity, Eq.~(\ref{eq:rhoS}) is in very good agreement with the exact  numerical solution obtained by integration of the dynamics Eq.~(\ref{eq:model}), as can be seen in Fig.~\ref{fig:revealed}. 
Notice the two different profiles, depending on the diffusion coefficient $D$: 
one gently following the landscape heterogeneity  and the other strongly oscillatory.

For small $D$, 
the induced oscillations display two evident characteristics, 
which depend on $\tilde{\gamma}_q$: 
a well-defined wavenumber and 
an amplitude that decays with the distance from  the interface 
at $x=\pm L/2$ (highlighted by dashed vertical lines in Fig.~\ref{fig:revealed}). 
We will see in the next section how the 
characteristics of the oscillations reflect the details of the kernel $\gamma_q$.

\subsection{Semi-infinite habitat}
\label{sec:semiinfinite}

Since oscillations are induced by changes in the landscape, 
it is worth focusing, from now on, on one of the interfaces between a more viable region  and a less viable one. 
Moreover, we assume a refuge much larger than the oscillations wavelength, 
sufficient to follow over several cycles the structure originated at the interface. 
To do that, we consider a semi-infinite habitat defined by
\begin{equation}
\label{eq:semiinfinite}
\psi(x) =   -A\Theta(-x) \,,
\end{equation}
where  for convenience the interface was shifted to $x=0$, such that the low-quality 
region is at $x<0$. 
As an additional feature, we consider that the harmful conditions are very strong, 
that is, $A \to \infty$. The purpose is twofold, on the one hand, it allows to test the 
robustness of the results beyond the small-$A$ approximation, on the other, it 
allows a simplification as follows. 
When $A \gg a$,   $\rho$ is very small in the unfavorable region, 
then the nonlinear competition term in Eq.~(\ref{eq:model}) can be neglected, 
leading to a steady distribution that decays exponentially from the interface as 
$\rho(x<0) \sim \exp[\sqrt{(A-a)/D}\,x]$. 
Thus,  in the limit $A\to \infty$,  we have $\rho(x <0,t)= 0$.  
In addition, the semi-infinite habitat is simulated by the interval $[0,L]$, where $L$ ($=100$ in our simulations) is large enough in comparison to oscillation length-scales.
Then, 
far away from the interface, we set $\rho(x \ge L,t)=\rho_0$.   
This is the setting used to produce Fig.~\ref{fig:profiles}b, 
by numerical integration of Eq.~(\ref{eq:model}).

\begin{figure}[h!]
\begin{center}
\includegraphics[width=0.5\columnwidth]{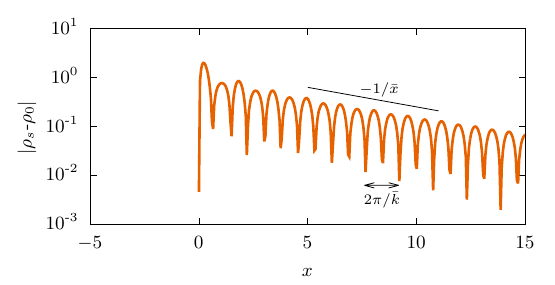}
\vspace{-0.2cm}
\end{center}
\caption{\textbf{Characterization of stationary profiles.}
Long-time solutions  approach a stationary state 
characterized by the wavelength $2\pi/\bar{k}$ and decay length $\bar{x}$. 
The slope of the straight black line is indicated in the figure.
This example was obtained from numerical integration of Eq.~(\ref{eq:model}), 
assuming a semi-infinite habitat, with
 parameters $D=  0.003$,  $\gamma_q(x)$ with $\ell=2$ and $q=-0.5$. 
}
\label{fig:measures}
\end{figure}

As sketched in Fig.~\ref{fig:measures}, for each steady distribution attained at long times, 
we measure the wavelength, from which we obtain the wavenumber $\bar{k}$,  
and the decay length $\bar{x}$, by observing that the envelope of the oscillations 
decays as $\exp(-x/\bar{x})$.

\begin{figure}[h!]
\begin{center}
\includegraphics[width=0.8\columnwidth]{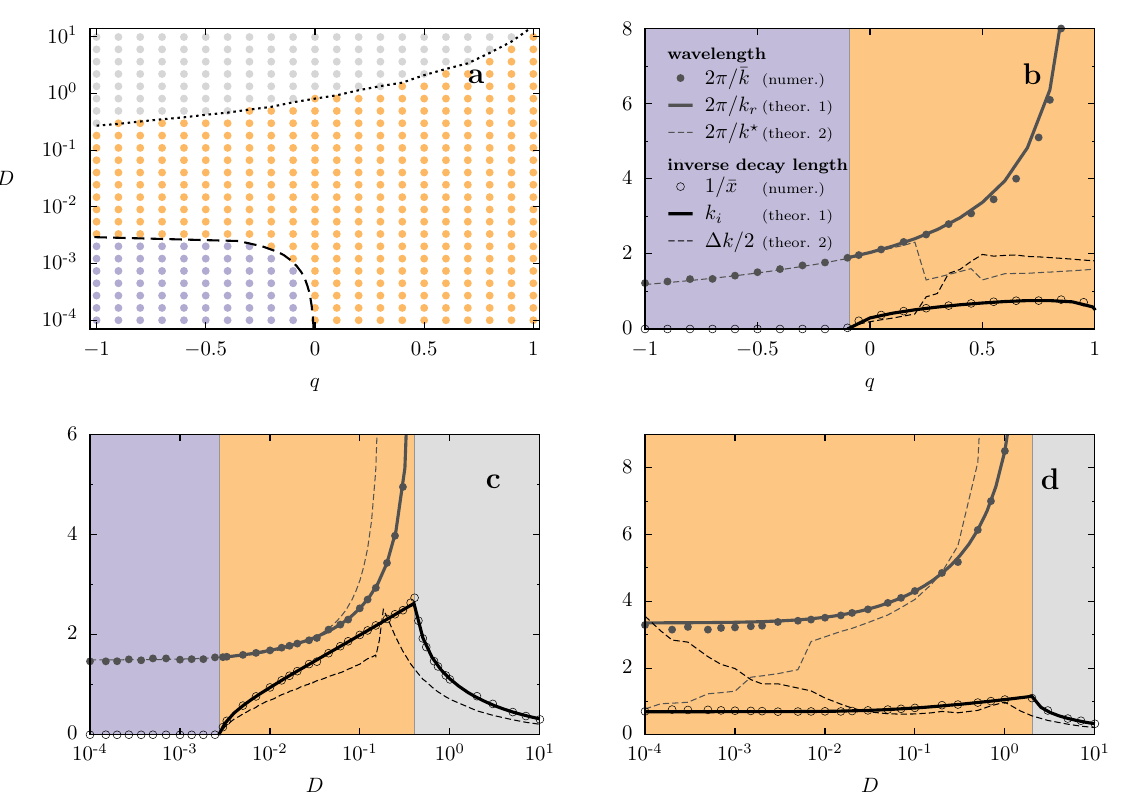}
\end{center}
\caption{ {\bf Phase diagram and characteristics of the stationary profiles as a function of diffusion coefficient $D$ and $q$,}
in the semi-infinite habitat. We used the kernel $\gamma_q(x)$, with $\ell=2$. 
(a) Phase diagram in the $q-D$ plane, and cuts at 
(b) $D=10^{-3}$, (c) $q=-0.5$ (d) $q=0.5$. 
The remaining parameters are $a=b=1$.
In diagram (a), for each point in the grid,  the type of regime was determined based on  
the values of $2\pi/\bar{k}$ and $\bar{x}$ that characterize the solutions of 
Eq.~(\ref{eq:model}): 
{\it sustained oscillations} ($\bar{k}>0$ and $\bar{x}\to \infty$, lilac), 
{\it decaying oscillations} ($\bar{k}>0$ and finite $\bar{x}$, orange), 
and {\it pure exponential decay} ($\bar{k}=0$ and finite $\bar{x}$, gray).
In (a), the dashed and dotted lines correspond to $k_i=0$ and $k_r=0$,  respectively, where $k_r$ and $k_i$ are the real and imaginary parts of the zeros of $\lambda (k)$, with the smallest positive imaginary part.  In (b)-(d), symbols correspond to measurements of numerical profiles, according to Fig.~\ref{fig:measures}, and solid lines correspond to the prediction in Eq.~(\ref{eq:prediction}) (theoretical 1).
Thin dashed lines correspond to 
the harmonic estimate (theoretical 2) given by Eq.~(\ref{eq:theory}).
}
\label{fig:diagrams}
\end{figure}

The stationary spatial structures that emerge for $x>0$ can be classified into the three types  depicted in Fig.~\ref{fig:profiles}:
{\it sustained oscillations} (lilac line, with $\bar{k}>0$ and $\bar{x}\to \infty$); 
{\it decaying oscillations}  (orange line, with $\bar{k}>0$ and finite $\bar{x}$); 
{\it exponential decay}   (gray line $\bar{k}=0$ and finite $\bar{x}$).
In the case of Fig.~\ref{fig:profiles}b, these three types appear when $D$ changes. 
We  also systematically varied the shape parameter $q$ to construct the 
phase diagram in the plane $q-D$ presented in Fig.~\ref{fig:diagrams}a.

To perform a theoretical prediction  of $\bar{k}$ and  $\bar{x}$, 
within the linear approximation, we consider 
that these oscillation parameters should be related to the poles of the integrand 
${\rm e}^{ikx} \tilde{\psi}(k) /[-\lambda(k)]$ in the expression for the 
inverse Fourier transform that provides the solution, according to Eq.~(\ref{eq:rhoS}). 
As far as  the external field $\psi(x)$ does not introduce non-trivial poles, like in the 
case of a Heaviside step function  ($\tilde\psi(k) \sim  1/k$), 
only the  zeros of the complex extension of $\lambda(k)$ matter. 
The dominant (more slowly decaying mode) is given by the complex poles 
$k=\pm k_r + i k_i$ ($k_r>0$) with minimal 
positive imaginary part that, except for amplitude and phase constants, will  
approximately provide patterns of the form ${\rm e}^{-k_i x} \cos (k_r x)$, allowing the identifications 
\begin{equation}
\label{eq:prediction}
\bar{k} = k_r\quad\text{ and }\quad 1/\bar{x}= k_i.  
\end{equation}
%
This theoretical prediction~\cite{poles} is in very good agreement with the 
results of numerical simulations,  as shown in Fig.~\ref{fig:diagrams}, explaining the observed regimes. 

Moreover,  the modes that persist beyond the interface have relatively small amplitudes, 
so that  the system response is approximately linear in this region.    

Lastly, recall that this analysis assumes mode stability ($\lambda(k)<0$). 
When $\lambda(k^\star)>0$, the system is intrinsically unstable, 
with the poles having null imaginary part (lying on the real axis). 
Nevertheless, the initially fastest growing mode, given by 
the maximum of $\lambda(k)$, tends to remain the dominant one in the 
long term~\cite{colombo12},   yielding $\bar k \simeq k^\star$ 
for the sustained oscillations ($\bar{x} \to \infty$).

In order to obtain  further insights, it is useful to consider the response function $\tilde R(k)$ 
that, from Eq.~(\ref{eq:eqepsilonFT}), is
\begin{equation}
\tilde R(k) \equiv \frac{|\tilde \varepsilon_s(k)|^2}{|\tilde{\psi}(k)|^2} = 
\frac{\rho_0^2}{\lambda^{2}(k)} \,.
\label{eq:spectrum}
\end{equation}
Despite missing some of the dynamical information contained in the phase of $\lambda(k)$, 
it can provide  a more direct estimation of the observed parameters  than through calculation of the poles. 
%
%
In order to perform this estimation, we resort to the response function of a driven damped linear oscillator~\cite{butikov2004} described by the equation
$\varepsilon_H''(x)+2\zeta
k_0\varepsilon_H'(x)+k_0^2\varepsilon_H(x)= f(x)$. We have
\begin{equation}
\tilde R_H(k) \equiv  \frac{|\tilde \varepsilon_H(k)|^2}{|\tilde{f}(k)|^2}=
\frac{1}{|\lambda_H(k)|^{2}} =
 \frac{1}{(k^2-k_0^2)^2+4\zeta^2k_0^2k^2}\, ,
\label{eq:RH}
\end{equation}
with $-\lambda_H(k) = -k^2   + i2\zeta k_0 k    + k_0^2$,
whose zeros (poles of $1/\lambda_H(k)$)
are $k= \pm k_r + i k_i = k_0 (\pm \sqrt{1-\zeta^2}+i\zeta )$,
where $k_0$ is the natural mode and $\zeta$ is the damping coefficient.
Note that, under a step forcing $f(x)=k_0^2 \Theta(x)$,
which simulates our present setting,
those poles carry the essential information of the damped-oscillation solution,
given by $ \tilde{\varepsilon}_H(k) = \tilde{f}(k)/[-\lambda_H(k)]$, 
where $\tilde{f}(k)=k_0^2(\pi\delta(k) -i/k)$.
In the underdamped case ($\zeta<1$), this solution is
explicitly given by
\begin{equation}
\varepsilon_H(x) = \left[1 - \frac{k_0}{\kappa} e^{-x/\xi} \sin(\kappa x  +\phi)\right]\Theta(x) \, ,
\label{eq:harmonicsolution}
\end{equation}
where $\kappa = k_0\sqrt{1 -\zeta^2}$ ($=k_r$), $\xi = 1/(\zeta k_0)$ ($=1/k_i$),
and the phase  constant $\phi=\tan^{-1}(\xi\kappa)$.
The solution for the overdamped case emerges for $\zeta>1$, when the
zeros of $\lambda(k)$ are pure imaginary with $k_i=k_0(\zeta \pm\sqrt{\zeta^2-1})$.
The connection between the poles of $\tilde R_H(k)$ and the
dynamic solution of the driven harmonic oscillator is possible
because, as previously discussed, $\tilde f$ does not introduce
relevant poles, and the forced solution has a form similar to
the unforced one.

The harmonic model is, in fact, the minimal model sharing
characteristics with our observed structures, and the
correspondence between Eqs.~(\ref{eq:RH}) and
~(\ref{eq:harmonicsolution}) will allow to estimate the
oscillation features. In the limit of small $\zeta$, $\tilde
R_H(k)$ has a sharp peak, characterized by a large quality
factor $Q \equiv k^\star/\Delta k$, where $\Delta k$ is the
bandwidth at half-height of $\tilde{R}(k)$ around
$k^\star$~\cite{butikov2004}.
First, we see that  the position of the peak of $\tilde R_H$
approximately gives the oscillation mode $\kappa$, according to
$k^\star = k_0\sqrt{1-2\zeta^2} = \kappa + {\cal O}(\zeta^2)$.
Second, the bandwidth is related to the decay-length through
$\Delta k = 2/\bar{x} + {\cal O}(\zeta^2)$~\cite{half-note}.

Putting all together, as long as  $\tilde{R}(k)$ resembles the
bell-shaped form of $\tilde{R}_H(k)$, we can use the following
estimates, which are correct for the harmonic case to first
order in $\zeta$:
\begin{align}
\bar{k} \simeq   \underset{k}{\text{arg max}}\,(\tilde{R}) \equiv k^\star
\quad\text{ and }\quad \bar{x} \simeq \frac{2}{\Delta  k} \,.
\label{eq:theory}
\end{align}
The expression for $\bar{x}$ is also valid in the overdamped
limit (large $\zeta$ in the harmonic model), in which case the
maximum is located at $k^\star=0$.

\begin{figure}[h!]
\begin{center}
\includegraphics[width=0.5\columnwidth]{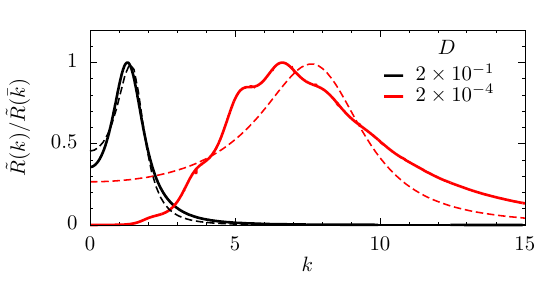}
\end{center}
\vspace{-0.2cm}
\caption{{\bf Comparison of $\tilde R(k)$ with the harmonic  response $\tilde R_H(k)$, 
both normalized to their maximal values.}
$\tilde R(k)$ of our model, given by Eq.~(\ref{eq:spectrum})  (solid lines) 
and harmonic response $\tilde{R}_H(k)$, given by Eq.~(\ref{eq:RH}) (dashed lines),  
where the values of $k_0$ and $\zeta$ were obtained by fitting Eq.~(\ref{eq:RH}) 
to $\tilde R(k)$.  
In all cases, $q=0.5$, $\ell=2$ and two different values of $D$ shown in the legend were considered. 
Notice that for $D=2\times 10^{-1}$, the response can be described by the harmonic approximation. 
For $D=2\times 10^{-4}$, the response is multipeaked, indicating 
that the harmonic approximation fails. 
In fact, the dominant mode observed in the simulations is not given by the absolute maximum, but by the small hump at 
$k\simeq 2.1$, as predicted by the analysis of complex poles. 
}
\label{fig:harmonic}
\end{figure}

The adequacy of the harmonic framework as an approximation to
the response function of our model, $\tilde{R}(k)$,   is
illustrated in Fig.~\ref{fig:harmonic}. 
In  the case $D=2\times 10^{-1}$, the harmonic response is able to emulate $\tilde R(k)$. 
Then, if the harmonic approximation holds, one expects that the 
estimates given by Eq.~(\ref{eq:theory}) should work for the population dynamics case.  
In fact, they do work, as we will see below. 
Differently, when $D=2\times 10^{-4}$, $\tilde R(k)$ does not follow the  
harmonic shape, 
it is multipeaked and the dominant mode observed in the
simulations is not given by the absolute maximum.

In Fig.~\ref{fig:diagrams}, we compare the values of $\bar{k}$ and $\bar{x}$ 
extracted from the numerical solutions of Eq.~(\ref{eq:model}) with those 
estimated by Eq.~(\ref{eq:theory}) (dashed lines) 
and, more accurately, with those predicted 
from the poles of $\tilde{R}(k)$  (solid lines), which perfectly follow 
the numerical results. 
The harmonic estimates are shown in the full abscissa ranges, as a reference, even in regions where 
the approximation is not expected to hold, because discrepancies give an idea of the departure from the harmonic response.

Figure~\ref{fig:diagrams}c shows outcomes 
for a fixed $q<0$ ($q=-0.5$), corresponding to a vertical cut in the diagram of Fig.~\ref{fig:diagrams}a. 
Sustained oscillations (i.e., $\bar{x} \to 0$) can emerge for $q<0$, when diffusion is weak, 
namely, for $D<D_c \simeq 0.0025$ (lilac colored region), 
where $D_c$ is obtained from $\lambda(k^\star)=0$. 
When $D$ increases beyond this critical value, 
oscillations are damped with a finite characteristic length $\bar{x}$. 
For even larger values of $D$, 
oscillations completely disappear ($\bar{k}\to 0$).  
Note that the comparison between numerics and harmonic theory (symbols vs. dashed lines) is good close to 
the pattern transition point $D_c$, where the response peak is sharp (large $Q$). 
Despite the lack of agreement for larger $D$, the harmonic approximation qualitatively works 
with a shift of the transition from attenuated oscillations to exponential decay. 

Figure~\ref{fig:diagrams}d (which corresponds to vertical cut 
at $q=0.5$ in the diagram of Fig.~\ref{fig:diagrams}a) 
shows the corresponding results for a fixed $q>0$ ($q=0.5$), which is
 characterized by the absence of sustained patterns. 
Above $D\simeq 0.02$, the response $\tilde{R}(k)$ is unimodal,  
a bell-shaped curve that resembles the harmonic response, as in the case $D=0.2$ (black lines) in Fig.~\ref{fig:harmonic}, 
producing a good agreement between harmonic and numerical results, despite being far from the large-$Q$ limit. 
However, for smaller values of $D$, the profile is multi-peaked, 
and not even $k^\star$ predicts the observed mode, indicating that the harmonic approximation does not hold, as for 
$D=2\times 10^4$ (red lines) in Fig.~\ref{fig:harmonic}.
In this regime, it is crucial to analyze the response function in terms of complex poles in order to extract the dominant mode and its decay.

Figure~\ref{fig:diagrams}b displays $\bar{k}$ and $\bar{x}$ as a function of $q$, 
for a fixed value of the diffusion coefficient ($D=10^{-3}$), 
corresponding to a horizontal cut in Fig.~\ref{fig:diagrams}a.  
Recall that, the smaller the value of $q$, 
the more confined is the interaction (thus, the larger is $\bar x$).  
For $q < q_c \approx -0.093$ there are sustained oscillations ($\bar x \to \infty$).  
Above $q_c$, oscillations decay, 
which is indicated by the transition of $1/\bar{x}$  from null to finite values. 
Again, near this transition, the harmonic approximation works well, 
but, far from the critical point, it fails,
as noticed above $q\simeq 0.2$, where there is a strong mismatch 
between the main mode given by the harmonic approximation and the numerical one. 
Also in this case, a small hump in the response function represents the dominant mode, as predicted by the analysis of the complex poles of $\tilde{R}$.

\subsection{Two-dimensional landscapes}
\label{sec:2D}

In this section, we show results of simulations for relevant 2D scenarios, verifying 
that the picture of induced oscillations described up to now for 1D  also holds in 2D. 

\begin{figure}[h]
\begin{center}
\includegraphics[width=0.8\columnwidth]{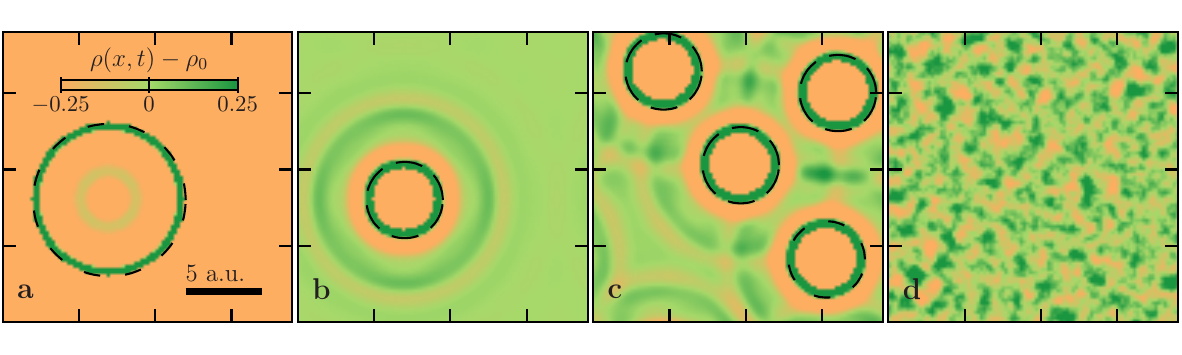}
\end{center}
\caption{ 
{\bf Long-time spatial distribution in 2D.}
Simulated scenarios: (a) a circular region (with radius 5 a.u., highlighted with a black dashed boundary) 
where the growth rate  is positive, $a$ (in a strong negative background $a-A$); 
(b) a circular region (with radius 2.5 a.u., highlighted with a black dashed boundary) 
where the growth is strongly negative $a-A$ (while outside, it is positive, $a$);
(c) four regions with negative growth rates $a-A$ (in a positive background, $a$);
(d) time-independent random landscape 
(where each spatial cell is assigned a growth rate uniformly distributed  in $[0.5a,1.5a]$).  
In all cases the interaction kernel is $\gamma_q$, with $\ell=2$ and $q=0.5$,  
 $D=10^{-3}$, $a=b=1$ and $A=10$~. 
Colors show the deviation from the homogeneous state $\rho(x,t)-\rho_0$ (where $\rho_0 = 1$ 
for the chosen values of the parameters). 
For numerical integration, a pseudo-spectral method~\cite{montagne}  was used with 
$\Delta x = 0.2$ and $\Delta t = 10^{-3}$.}
\label{fig:multi}
\end{figure}

Snapshots of simulations for different 2D landscapes are presented in Fig.~\ref{fig:multi}: 
 a refuge (a), a defect (b), multiple defects (c) and spatial randomness (d) 
where many spatial scales are present. 
It is worth remarking that, in 2D, for the kernel $\gamma_q$, patterns only appear in homogeneous landscapes if $q<q_c \simeq 0.25$ (i.e., if $\lambda(k^\star)>0$). 
Thus, in all the cases of Fig.~\ref{fig:multi} (using $q=0.5$)
we would not find oscillations if the landscape were homogeneous. 
In Fig.~\ref{fig:multi}, we see that for 2D the same picture as in 1D is found: decaying oscillations appear near landscape disturbances with a clear wavenumber and decay length. The linear response approach presented in Sec.~\ref{sec:semiinfinite}
can straightforwardly be extended to 2D.
Figure~\ref{fig:multi}a-c shows the case in which defects either increase or decrease the population growth rate. This can be promoted by ecosystem engineers such as termites~\cite{tarnita2017}.
Figure~\ref{fig:multi}d  shows a case where the landscape is random (in space, but time-independent). 
This situation, investigated in many previous studies~\cite{ridolfiBook,Sasaki1997}, 
produces a pattern  that is  noisy but has a dominant wavelength, which is related to $\ell$. 
Furthermore, although  there is not a clear identification of decay length from pattern observation, 
the linear theory would  allow one to estimate the characteristic spatial correlation length from the width of the Fourier spectrum.


\section{Inferring information about the interactions}
\label{sec:inferring}

\begin{figure}[h!]
\begin{center}
\includegraphics[width=1.0\columnwidth]{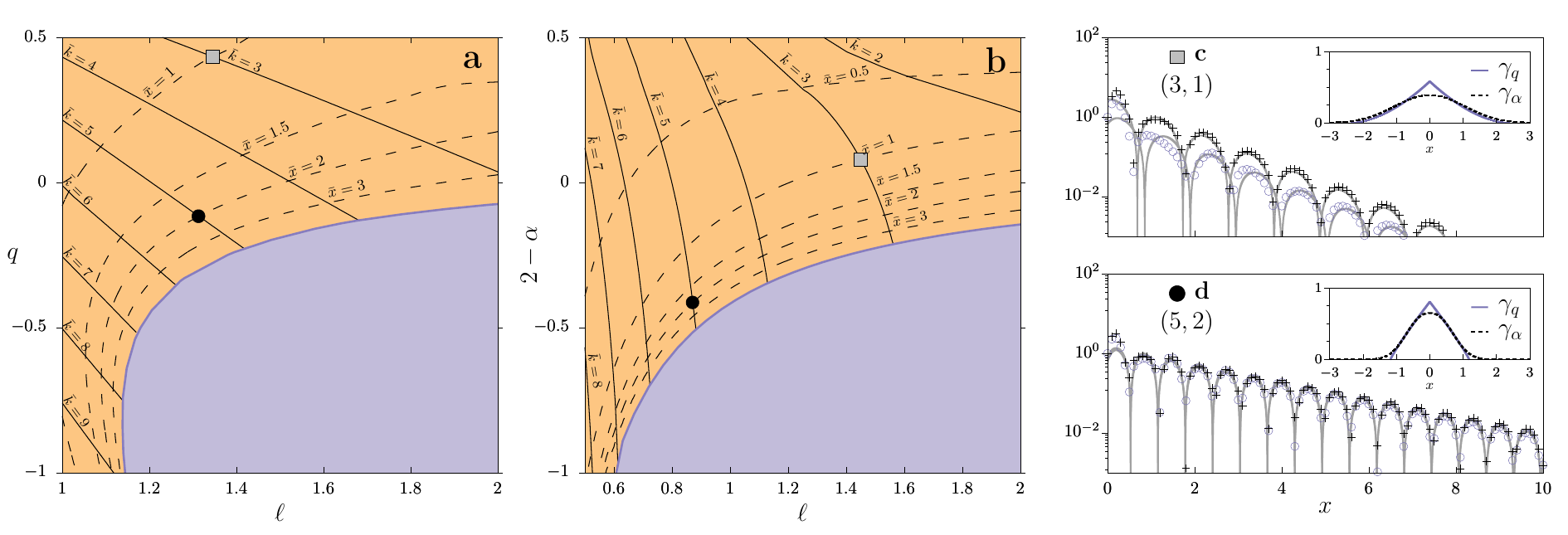}
\end{center}
\caption{ 
{\bf Determination of oscillation wavenumber, $\bar{k}$, 
and decay length, $\bar{x}$.} 
Contour lines for fixed wavenumber (solid lines) and decay lengths (dashed lines). Colors for different oscillatory regimes are applied to the background as in previous figures.
We considered the interaction kernel $\gamma_q$, given by Eq.~(\ref{eq:kernel}), in (a) and $\gamma_\alpha\equiv N_\alpha^{-1}e^{-|x/\ell|^\alpha}$ in (b). 
The remaining parameters are $D=10^{-3}$ and $a=b=1$. 
The highlighted points correspond to $(\bar{k},\bar{x})=(3,1)$ (gray square) and $(\bar{k},\bar{x})=(5,2)$ (black circle). 
(c)-(d) Oscillations produced by the kernels shown in the respective insets (case $\gamma_q$ in purple circles and $\gamma_\alpha$ in black crosses). 
For both panels, the inset shows the kernels $\gamma_q$ (solid purple) and $\gamma_\alpha$ (dashed black) obtained from the oscillation's parameters for the highlighted points: 
(c) $(\ell,q) = (1.345,0.433)$, $(\ell,2-\alpha) = (1.450,0.079)$ (gray squares), and  
(d) $(\ell,q) = (1.313, -0.116)$, $(\ell,2-\alpha) = (0.871,-0.413)$ (black circles). 
The gray lines show a sinusoidal fit for each case, namely $\rho_H(x) = 1 + Be^{-x/\bar{x}}\sin(\bar{k}x + \phi)$, with mode, $\bar{k}$, and decay, $\bar{x}$, as predicted by the mapping, while $B$ and $\phi$ were adjusted.
}
\label{fig:nullclines}
\end{figure}

In this section, we extend the discussion about the mapping between  kernel and oscillation parameters, showing how 
 information about the interactions can be extracted from landscape-induced oscillations.
For that purpose, using the theoretical predictions  given by Eq.~(\ref{eq:prediction}), 
we  obtained the contour lines for certain wavelengths 
$\bar{k}$ and decay lengths $\bar{x}$, in the space of the 
kernel parameters, 
as shown in the plane  $(q,\ell)$ of  Fig.~\ref{fig:nullclines}a, 
for the kernel 
$\gamma_q$:   $\bar{k}(\ell,q) = \text{constant}$, and $\bar{x}(\ell,q) = \text{constant}$.

These contour lines depend both on  $\ell$ and $q$. However,  while $\bar{k}$ is strongly controlled by the interaction scale, $\ell$,   $\bar{x}$ is more closely related to the shape parameter $q$.  As a consequence, there is a crossing of the lines that  uniquely  identifies the kernel properties. 
Of course, this is possible for the decaying-oscillation phase (orange region), in which oscillations have a well-defined $\bar{k}$ and $\bar{x}$. For the sustained-oscillation ($\bar{x}\to\infty$) and the exponential relaxation ($\bar{k}=0$) phases, the stationary distribution does not carry sufficient information to infer the specific values of $q$ and $\ell$ (in the perspective of the linear theory).

For comparison, in Fig.~\ref{fig:nullclines}b,  we perform the same analysis considering the stretched-exponential kernel, 
$\gamma_\alpha(x) \equiv N_\alpha^{-1} \exp(-|x/\ell|^\alpha)$, 
where $N_\alpha$ is the normalization factor. 
Likewise $\gamma_q$, this kernel also allows us to contemplate changes in the range and shape of the competitive interactions. 
Since, for $\gamma_\alpha$, compactness increases with $\alpha$, and sustained spatial oscillations require  $2-\alpha<0$~\cite{pigolotti}, in Fig.~\ref{fig:nullclines}b, we show contour lines for the same values used in Fig.~\ref{fig:nullclines}a,
in the plane defined by parameters  $2-\alpha$ and $\ell$. 
Also, in this case, the results show that the identification of the kernel features is possible. Additional results using this kernel can be found in the {\it Supplementary Information}.

To test the inference procedure, we imagine the scenario in which spatial oscillations with certain values of $\bar{k}$ and $\bar{x}$ are observed. Then, assuming that the population distribution evolves according to Eq.~(\ref{eq:model}), one proposes a generalized form for $\gamma$ (as one of the two discussed above) and extracts the kernel parameters from the $(\ell,\beta) \leftrightarrow (\bar{k},\bar{x})$ mapping (Figs.~\ref{fig:nullclines}a-b), 
where $\beta$ represents either $q$ or $2-\alpha$ (for $\gamma_q$ and $\gamma_\alpha$, respectively).

In Fig.~\ref{fig:nullclines}c-d, we verify that, in fact, using the extracted kernel parameters, $(\ell,\beta)$, numerical simulations produce spatial oscillation with the correspondent $(\bar{k},\bar{x})$ (black circles and gray squares).  
Moreover, in the insets of Fig.~\ref{fig:nullclines}c-d, we compare the inferred kernel under the $\gamma_q$ and $\gamma_\alpha$ representations. Note that, regardless of the particular choice made for $\gamma$, both profiles have the same coarse-grained appearance, allowing to qualitatively access the characteristic length and compactness of the influence function. 
However, we stress that since the information provided by the theory is limited, it is not possible 
to infer exactly the form of $\gamma$ just by measuring $(\bar{k}, \bar{x})$ of the oscillations. 
To access the fine details about the kernel, improvements of this methodology could look for information encoded in the spatial transient and nonlinear effects that occur close to the interface.


\section{Final remarks}
\label{sec:final}

Heterogeneities can modify system stability conditions~\cite{ridolfiBook,garcia2012noise,horsthemke1984noise}, inducing the emergence of states
that would not be present under homogeneous conditions. In the
context of biological populations with the potential to develop
spatial patterns, it would be interesting to establish if the
conditions for the occurrence of pattern formation can be
modified by the presence of environmental inhomogeneities.
Also, it is natural to ask under which conditions or how
heterogeneities can be used to help in the task of identifying
details about microscopic interaction from the observation of
the macroscopic patterns~\cite{zhao2020}.

The first question was considered by Page et al.~\cite{page2003} in the context of two-species reaction-diffusion models
undergoing a pattern formation instability of the Turing type.
It was found that the range of parameters for which periodic
solutions were possible was extended by the presence of a
discontinuity in some system parameter. Here, we have addressed
this issue in a model of a population of competing
organisms. In particular, we have considered a nonlocal FKPP
equation which includes reproduction, diffusion, and
competition between individuals at a distance. Non-local
competitive interactions can arise due to different mechanisms,
as in root-mediated competition for water in
vegetation~\cite{ricardogrl,tarnita2017,oto2013},   
and release or consumption of intermediate substances by the organisms~\cite{ricardogrl,ricardo2014,ojalvo2017,bauerle2018}.
In all cases, the fact that competitive interactions are
nonlocal plays a major role in the spatial organization of the
population. In particular, pattern formation can occur in a
manner related to the Turing case, although in the FKPP
approach a single species is explicitly modeled, with competitive
interactions effectively captured by the influence
function, $\gamma$. The kernel family $\gamma_q$ was chosen in
the examples because it allows shapes of different compactness.
But results for another important class, the stretched
exponential family, were presented in {\it Supplementary Information}. A necessary condition
for the development of stable spatially periodic patterns
starting from a uniform solution is that the interaction
profile is sufficiently compact, meaning sub-triangular ($q <
0$), for the $q$-exponential (see Fig.~\ref{fig:kernel}a), or platykurtic
($\alpha > 2$), for the stretched exponential family (see Fig.
S1).

Previous work already showed that these conditions become less
strict when the initial condition contains sharp changes:
propagating-front solutions of the nonlocal FKPP equation
develop oscillatory patterns in cases when the influence
function does not satisfy the above compactness
condition~\cite{Sasaki1997,kessler18}. Here, we have
investigated how the above scenarios are modified  by an abrupt
change in the ecological landscape. We have seen that modes
which are suppressed in the homogeneous-landscape case are
activated by the interface, producing decaying oscillations.
Activation of modes have been observed in particular
realizations of the nonlocal FKPP triggered by random~\cite{Sasaki1997,silva2014} heterogeneities. In these cases, the
variation of the equation parameters extends to all the space,
generating a noisy pattern that has a clear dominant wavelength
(see Fig.~\ref{fig:multi}d). A localized perturbation (like the interface we
consider) is more useful because it puts into evidence, besides
the dominant wavenumber, also the decay length (see Fig.~\ref{fig:multi}b),
which reflects how a perturbation  spreads in the system. When
the landscape variation occurs in all points of space, the
decay length is blurred. The implication of these results
(alike in a  multispecies reaction-diffusion situation~\cite{page2003}) 
is that the range of parameters
for which spatial structures can occur in biological
populations can be much larger than superficially expected.

Deepening further into the interplay between nonlocality and
environmental heterogeneity, we have shown here that the
presence of heterogeneities can reveal information about
interaction scales that would otherwise be hidden. This is
possible due to the existence, once a functional form for the
influence function is fixed, of a one-to-one correspondence
between the parameters of the competitive interactions (shape
and range) and the landscape-induced oscillation  features
(wavenumber and decay length). So, the natural or artificial
interposition of an interface  can act as a lens that allows us
to see what is veiled in a homogeneous landscape. This might be
particularly useful in situations where the details of the
long-range influence are not perfectly clear, depending on a
complex combination of ecological factors. Although the
approach provides the coarse-grained profile of the influence
function, without distinguishing other fine details, our
results allow to take a step forward  in the direction of
understanding the connection between interactions at the
individual level and the emergent macroscopic patterns~\cite{zhao2020}.

The numerical results were perfectly predicted by the analysis of the poles of the system response function. Additionally, we have presented a harmonic approximation that has limitations but provides a more direct insight.  
The analytical results were obtained for
general forms of the landscape $\Psi(x)$ and can be used to
understand the effects of arbitrary  heterogeneous landscapes,
like the multiple and random cases shown in Fig.~\ref{fig:multi}. But we have focused on the case of a single interface 
because of its above-discussed features. 

It is worth remarking that,
due to computational cost, we compared theoretical predictions
with numerical simulations mostly for 1D, but we showed also
similar outcomes in some 2D environments. Lastly, it is also
interesting to remark that the reported results can reach
contexts beyond population dynamics, since the  interplay
between nonlocality and heterogeneity is found in diverse
systems.

\vspace{2cm}

\section*{Acknowledgements}

E.H.C., C.L. and E.H.G. acknowledge financial
support from Agencia Estatal de Investigaci\'on through the Mar\'{\i}a de Maeztu
Program for Units of Excellence in R\&D (MDM- 2017-0711). 
V.D. and C.A. acknowledge partial financial support by the 
Coordena\c c\~ao de Aperfei\c coamento de Pessoal de N\'{\i}vel Superior
 - Brazil (CAPES) - Finance Code 001 and also by 
 Conselho Nacional de Desenvolvimento Cient\'{\i}fico
e Tecnol\'ogico (CNPq), and Funda\c c\~ao de Amparo \`a Pesquisa
do Estado do Rio de Janeiro (FAPERJ).

\section*{Author contributions statement}

V.D., E.H.C., C.L., E.H.G. and C.A. performed the research and wrote the manuscript.    
All authors reviewed the manuscript. 

\section*{Additional information}

\textbf{Accession codes} (not applicable); \\
\textbf{Competing interests} (The authors declare no competing financial interests).

\newpage

\appendix
\setcounter{figure}{0}
\setcounter{equation}{0}
\renewcommand{\theequation}{S\arabic{equation}}
\renewcommand{\thefigure}{S\arabic{figure}}

\section*{Supplementary Information}

\section*{Numerical method}
 
The numerical solution of  the one-dimensional generalized FKPP equation   presented in Eq.~(3) of the main text  was obtained by means of a  finite-difference  forward-time-centered-space (FTCS) scheme, implemented in C language. 

For the spatial grid, we consider  equally spaced mesh points
$x_i =i\Delta x$, with integer $i$ and grid space $\Delta x$.
At each mesh point $x_i$, the density at time $t_j$, 
$\rho_j^i=\rho(x_i,t_j)$, evolves according to an explicit scheme 
with fixed time step $\Delta t$, such that $t_j= j\Delta t$ with integer $j$.  
For the spatial discretization of the second derivative, we used   the centered form
$    \partial_{xx}\rho_j^i =( -\rho_j^{i+2}+16\rho_j^{i+1}-30\rho_j^i+16\rho_j^{i-1}-\rho_j^{i-2})/(12\Delta x^2)$, and for the integration of the nonlocal competition term we used the trapezoidal rule. Lastly, a fourth-order Runge-Kutta approximation for the new values $\rho_i^{j+1}$ was used~\cite{numerical}, which improves the stability domain in comparison with the Euler method.
Typical values used for  spatial and temporal discretization, respectively, are   
$\Delta x = 0.05$  and $\Delta t \in [10^{-2},10^{-4}]$ (depending on $D$), leading to an estimated relative error smaller than $10^{-4}$.

For the initial preparation of the system, we applied  a small random perturbation around the homogeneous solution $\rho_0=a/b$, 
drawing a  random number $\xi^i$ uniform in $(-\epsilon,\epsilon)$, with $\epsilon \ll \rho_0$, for each mesh point $x_i$, 
such that, $\rho^i_0=\rho_0+\xi^i$.

The boundary conditions for each configuration are described below 
(in Section \textit{Boundary conditions}) and in Section \textit{Stationarity condition}, we show the stop criterion used to determine stationarity.

The two-dimensional simulations in Fig. 7 of the main text were performed using the pseudospectral algorithm of Ref. [46], with $\Delta t=10^{-3}$ and $\Delta x=0.2$.

\subsection*{Boundary conditions}

Different boundary conditions were adopted along the main text:

\begin{figure}[b!]
\begin{center}
\includegraphics[width=0.5\columnwidth]{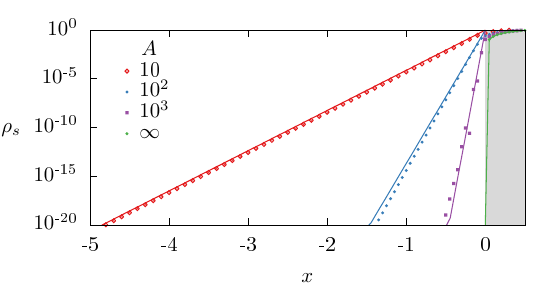}
\end{center}
\caption{ \textbf{Exponential decay from the interface (for $A>>a$).}
Numerical integration for the semi-infinite habitat with very strong harmful conditions outside the habitat (symbols), for different values of $A$ indicated in the legend.  The grey region (at $x>0$), represents the refuge 
near the interface.
The solid lines represent the exponential decay predicted by $\rho(x\le 0) \sim e^{\sqrt{(A-a)/D}\,x}$.
The influence kernel is $\gamma_q$ with $q=0.1$ and $\ell=2$, $D=0.1$ and $a=b=1$.
 }
\label{fig:expdecay}
\end{figure}

(i) For the homogeneous landscape (Fig.~1a) and for (finite) refuge case (Fig. 3), 
we used periodic boundary conditions.   

(ii) For the semi-infinite habitat (e.g., in Fig.~\ref{fig:expdecay}, with growth rate $a$ in the region $x\ge 0$, and
with growth rate $a-A$, otherwise), 
the integration was performed in a grid with $-L\le x \le L$, 
using $L=100$, much larger than oscillation  length-scales, under
the constraints $\rho(x \le -L)= (a-A)/b$ and $\rho(x \ge L)=\rho_0=a/b$.

(iii) For the semi-infinite habitat with strong harmful conditions, $A \to \infty$ (see profiles in Figs. 1b, 4, 8c, 8d),
the integration was performed in a grid with $0 \le x \le L$, 
using $L=100$, under the 
conditions $\rho(x < 0)= 0$ and 
$\rho(x\ge L)=\rho_0=a/b$. This choice is justified by the fact that 
in the limit $A\to \infty$, the density outside the refuge 
vanishes, as shown in  Fig.~\ref{fig:expdecay}.

\subsection*{Stationarity criterion}

\begin{figure}[h!]
\begin{center}
\includegraphics[width=0.8\columnwidth]{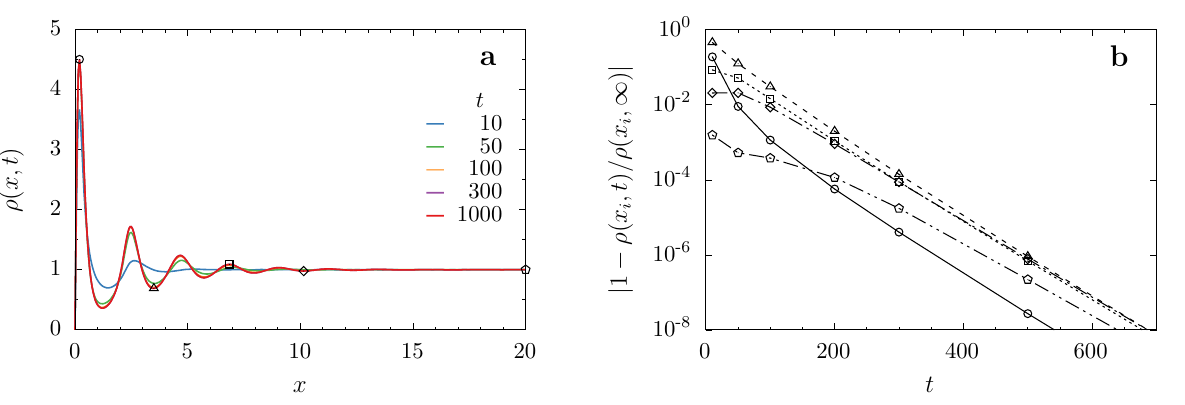}
\end{center}
\caption{ \textbf{Approaching the stationary profile.} (a) Decaying oscillations in population density in a semi-infinite refuge with strong harmful conditions outside ($A \to \infty$), at different times indicated in the figure. 
%
Eq.~(3) of the main text, using   $\gamma_q$  with $q=0.1$ and $\ell=2$, and  $D=10^{-3}$, was numerically integrated 
using $\Delta x=0.05$ and $\Delta t=0.01$. 
(b) Relative difference $|1- \rho(x_i,t)/\rho(x_i,\infty)|$ 
is plotted vs. time, for the values of $x_i$ selected in panel (a), 
identified by the same symbols, 
showing that stationary values are exponentially approached.
}
\label{fig:ev}
\end{figure}

Figure~\ref{fig:ev}a shows the population density $\rho(x,t)$ vs. $x$ at different instants of time $t$, 
obtained from the integration of Eq.~(3) for a semi-infinite refuge with strong harmful conditions outside ($A\to\infty$). 
As time passes, the profile progressively attains a stationary form, $\rho(x,\infty)$. This limiting value can be estimated by noting that the relative difference (discrepancy) between the profiles at different instants decays exponentially with time. 
In Fig.~\ref{fig:ev}b, the discrepancy
$|1- \rho(x_i,t)/\rho(x_i,\infty)|$, 
for selected mesh points $x_i$ is displayed, 
showing exponential convergence.
The final simulation time was chosen such that 
the discrepancy is smaller than $10^{-4}$ for all the mesh points 
in the interval of interest.

\section*{Results for the stretched exponential kernel}

We consider, as a second class of kernels, the stretched exponential family,  
\begin{equation}
\gamma_\alpha(x) = \frac{e^{-(|x|/\ell)^\alpha}}{2\ell \,\Gamma(1+1/\alpha)} \,,
\label{eq:kernelII}
\end{equation}
with $\alpha>0$ (to guarantee normalization). 
When $\alpha=1$, it produces the double exponential kernel, 
whose Fourier transform is 
$\tilde{\gamma}_1(k)=\frac{1}{1+ k^2\ell^2}. $ 
It includes the Gaussian ($\alpha=2$), whose Fourier transform 
is  
$\tilde{\gamma}_2(k)= e^{- k^2\ell^2/4}$. 
And it also reproduces the top-hat kernel in the limit $\alpha\to\infty$, that has $\tilde{\gamma}_\infty(k)= \sin(k\ell)/(k\ell)$.

\begin{figure}[h!]
\begin{center}
\includegraphics[width=\columnwidth]{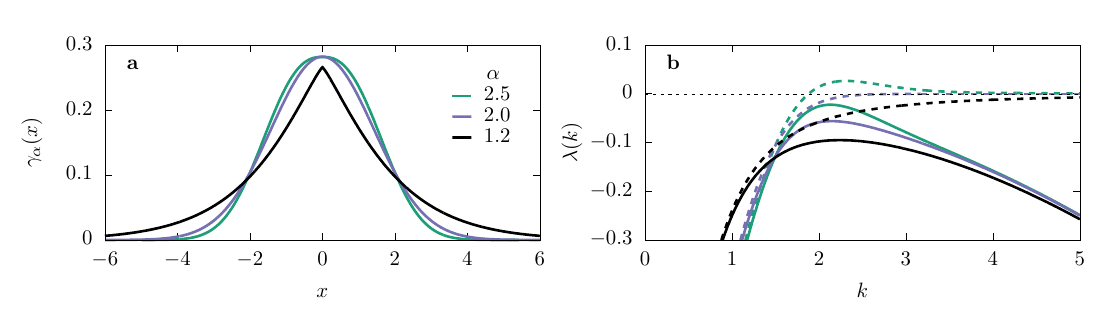}
\end{center}
\caption{ {\bf Interaction kernel and mode stability in a homogeneous medium.} 
(a) $\gamma_\alpha(x)$,  defined in Eq.~(\ref{eq:kernelII}), for the values of $\alpha$ indicated on the figure, and $\ell=2$.
(b) Mode growth rate $\lambda(k)$, for $a=b=1$,  
with $D=0$ (dashed lines) and $D=10^{-2}$ (solid lines), corresponding to 
the values of $\alpha$ plotted in (a). The case $\alpha=2$ is the critical one, 
for which the maximal value of $\lambda(k)$ at finite $k$ is zero when $D=0$.}
\label{fig:kernelII}
\end{figure}

\begin{figure}[h!]
\begin{center}
\includegraphics[width=\columnwidth]{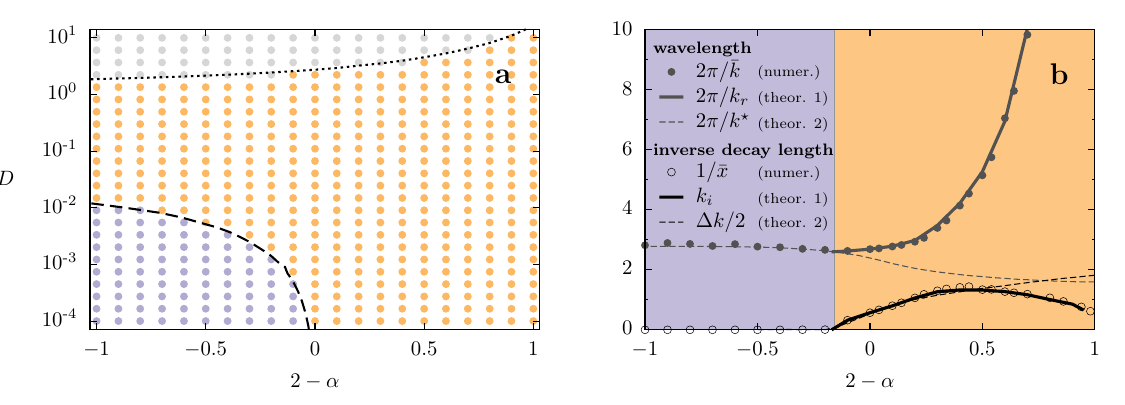}
\end{center}
\caption{ \textbf{Phase diagram  and characteristics of the stationary profiles as a function of shape parameter $2-\alpha$}, 
for kernel $\gamma_\alpha(x)$ with $\ell=2$.  
The remaining  conditions and conventions are as in Fig.~5 of the main text.
In panel b, $D=10^{-3}$.
 }
\label{fig:regimes}
\end{figure}

In Fig.~\ref{fig:kernelII}, we show the mode growth rate, $\lambda(k)$, for three values of $\alpha$. 
And in Fig.~\ref{fig:regimes}a, we show the phase diagram.
In Fig.~\ref{fig:regimes}b, we characterize the profiles as a function of parameter $2-\alpha$, for $D=10^{-3}$.  
All these results qualitatively resemble those produced in the main text for kernel $\gamma_q$ (Fig. 2 and 5 of the main text).

\begin{thebibliography}{10}
\bibitem{poles}
\bibinfo{note}{The zeros of the extension of $\lambda(k)$ to the complex plane were obtained numerically, by using
  the Taylor expansion of $\lambda(k)$ around $k=0$ and solving $Dk^2 +
  \sum_{n=0}^N \frac{1}{n!}\frac{d^n \tilde{\gamma} }{ dk^n}|_{k=0} \;k^n =0$,
  in the limit of sufficiently large $N$.}
\bibitem{half-note}
\bibinfo{note}{If $k_-<k_+$ are the points at which
  $\tilde{R}(k_\pm)=\tilde{R}(k^\star)/2$, then $\Delta k/2=(k_+-k_-)/2$. If
  only $k_+$ exists then we estimated $\Delta k/2=k_+-k^\star$.}
\end{thebibliography}
\end{document}